# Quantum heat transport in nonequilibrium anisotropic Dicke model[*]

KONG Junran,　MAO Mang,　LIU Huan,　WANG Chen

Department of Physics, Zhejiang Normal University, Jinhua 321004, China

**Abstract**

Nonequilibrium heat transport and quantum thermodynamics in light-matter interacting systems have received increasing attention. Quantum thermal devices, e.g., heat valve and head diode, have been realized. Recently, it has been discovered that the anisotropic light-matter interactions can greatly modify the eigenvalues and eigenvectors of hybrid quantum systems, leading to nontrivial quantum phase transitions, quantum metrology, and nonclassicality of photons. To explore the influences of anisotropic light-matter interactions on quantum transport, we investigate heat flow in the nonequilibrium anisotropic Dicke model. In this model, an ensemble of qubits collectively interacts with an anisotropic photon field. Each component interacts with bosonic thermal reservoirs. Quantum dressed master equation (DME) is included to properly study dissipative dynamics of the anisotropic Dicke model. Within the eigenbasis of the reduced anisotropic Dicke system, strong qubit-photon couplings can be properly handled. Our results demonstrate that anisotropic qubit-photon interactions are crucial for modulating steady-state heat flow. In particular, it is found that under strong coupling the heat flow is dramatically suppressed by a large anisotropic qubit-photon factor. While under moderate coupling, the anisotropic qubit-photon interactions enhance the heat flow. Moreover, the increase in the number of qubits amplifies the flow characteristics, with the peaks increasing and the valleys decreasing. Besides, we derive two analytical expressions of heat flows in thermodynamic limit approximation with limiting anisotropic factors. These heat currents exhibit the cotunneling heat transport pictures. They also serve as the upper boundaries for the heat flows in the finite-size anisotropic Dicke model. We also analyze the thermal rectification effect in the anisotropic Dicke model. It is found that a large temperature bias, a large anisotropic qubit-photon factor, and nonweak qubit-photon coupling are helpful in achieving the giant thermal rectification factor. We hope that these results can deepen the understanding of quantum heat transport in the anisotropic quantum light-matter interacting systems.

## 1. Introduction

Quantum optics is an interdisciplinary field combining atoms, molecules and photons, which mainly studies the interaction between photons and quantum matter[1,2]. A key area of research in quantum optics is quantum electrodynamics (QED)[3]. The research content of QED usually involves two-level qubit coupled single-mode light field. This quantum light-matter interaction opens up new directions for advancing quantum technology[4,5]. Solid-state QED device, as a new type of quantum device, has received close attention in the study of quantum thermal transport[6,7]. In cavity-QED devices, the coupling strength between the photon and the qubit is generally weak[3]. Such devices can be described theoretically by the Jaynes-Cummings (JC) model[8,9]. Recently, researchers have successfully implemented circuit-QED (cQED) systems using solid-state quantum engineering techniques. This kind of system can realize strong coupling, ultrastrong coupling and even deep strong coupling[10,11]. Typical solid-state cQED devices include quantum dot-cQED[12,13], superconducting-cQED[11], trapped-ion trap[14,15], etc. Such systems can usually be described by the quantum Rabi model (QRM). The QRM contains equally weighted rotating and anti-rotating wave terms[16–19]. At the same time, the properties of anisotropic QRM have been extensively studied. In contrast to the QRM, the rotating and anti-rotating wave terms in the anisotropic QRM Hamiltonian are unequally weighted[20]. It is shown that the anisotropic QRM can produce new features in quantum phase transition[20,21], quantum metrology[22,23] and photonic nonclassicality[24–26]. In addition, the intrinsic property of squeezed photon state of anisotropic QRM can also realize photon quadrature squeezing[27].

The Dicke model can be used to describe the interaction between multiple qubits and a single-mode quantum light field[28,29]. The Dicke model usually exhibits superradiance[30–33]. In the application of quantum batteries, it is found that the collective excitation of photons and qubits can significantly improve the charging performance[34–37]. In practice, quantum components such as photons and qubits inevitably interact with the external environment. This will lead to quantum dissipation[38]. The corresponding open Dicke model has a wide range of applications. Topics covered include nonreciprocal phase transition[39], multistable[40,41], discrete-time crystal[42,43], quantum laser[44,45], etc. The anisotropic Dicke model also has the characteristics of non-trivial phase transition[46,47], non-ergodic[48] and high-precision measurement[49].

Quantum heat transport and quantum thermodynamics in cQED have attracted great attention in recent years[7]. This kind of system usually consists of a superconducting qubit coupled to two optical harmonic oscillators. By adding two mesoscopic heat sources with temperature difference to the two harmonic oscillators, the steady-state heat flux can be observed. Therefore, the system can be fabricated into quantum thermal devices, such as thermal valve[6] and thermal diodes[50]. In addition, photonic thermal transport has been realized in a three-terminal superconducting circuit[51]. Researchers have also theoretically proposed several methods to realize the[54–56] of thermal rectification[52,53] and thermal amplification, among which the longitudinal photon-qubit coupling gives a new way to obtain significant thermal amplification effect[24]. Recently, Andolina et al.[57] studied the superradiant heat flow behavior in a multi-qubit cQED system. In this system, the photon field is eliminated and the quantum system is approximately represented as a collective qubit. Considering the above research status, a natural question arises: how does the anisotropic photon-qubit coupling affect the behavior of the quantum heat current in the nonequilibrium anisotropic Dicke model?

In this paper, we mainly study the characteristics of quantum heat transport in the anisotropic Dicke model. The dissipative dynamics of the anisotropic Dicke model is obtained by introducing the quantum dressed-state master equation (DME)[58–60]. This equation can reasonably deal with the photon-qubit strong coupling problem based on the intrinsic basis vectors of the anisotropic Dicke model. The structure of this paper is arranged as follows: Section 2 introduces the nonequilibrium anisotropic Dicke model and DME; Section 3 studies steady state heat flow and thermal rectification; Finally, Section 4 summarizes the results of the study.

**2. Nonequilibrium anisotropic Dicke model.**

In this section, the anisotropic Dicke model is first described, and the Hamiltonian of two coupled oscillators is approximated in the thermodynamic limit, and then the quantum master equation is introduced in the framework of dressed States of hybrid quantum systems.

**2.1 Model**

It is well known that the famous Dicke model consists of many qubits coupled with a single-mode photon field[29], and its Hamiltonian is

$$\hat{H}_{\text{Dicke}} = \varepsilon \hat{J}_z + \omega_a \hat{a}^\dagger \hat{a} + \frac{\lambda}{\sqrt{N_s}} [(\hat{a}^\dagger \hat{J}_- + \hat{a} \hat{J}_+) + (\hat{a}^\dagger \hat{J}_+ + \hat{a} \hat{J}_-)], \quad (1)$$

Where $\varepsilon$ is the Zeeman splitting energy; $\hat{J}_z = \frac{1}{2} \sum_{n=1}^{N_s} \hat{\sigma}_z^n$, $\hat{J}_+ = \sum_{n=1}^{N_s} \hat{\sigma}_+^n$ and $\hat{J}_- =$

$(\hat{J}_+)^\dagger$ the angular momentum operators describing the collective behavior of Ns qubits, where $\hat{\sigma}_+ = (\hat{\sigma}_x + i\hat{\sigma}_y)/2$, $\hat{\sigma}_x, \hat{\sigma}_y$ and $\hat{\sigma}_z$ are Pauli operators; $\hat{a}^\dagger(\hat{a})$ represents the creation (annihilation) of a photon of frequency $\omega_a$ in the radiation field; $\lambda$ is the coupling strength of the harmonic oscillator to the qubit. The Dicke model contains both a rotating wave term $(\hat{a}^\dagger \hat{J}_- + \hat{a}\hat{J}_+)$ and an anti-rotating wave term $(\hat{a}^\dagger \hat{J}_+ + \hat{a}\hat{J}_-)$. These two terms are equally weighted in the Dicke model. When considering the anisotropic Dicke model[20,48] shown in Fig. 1(a), the Hamiltonian is expressed as

$$\hat{H}_s = \varepsilon \hat{J}_z + \omega_a \hat{a}^\dagger \hat{a} + \frac{\lambda}{\sqrt{N_s}}[(\hat{a}^\dagger \hat{J}_- + \hat{a}\hat{J}_+) + \gamma(\hat{a}^\dagger \hat{J}_+ + \hat{a}\hat{J}_-)], \quad (2)$$

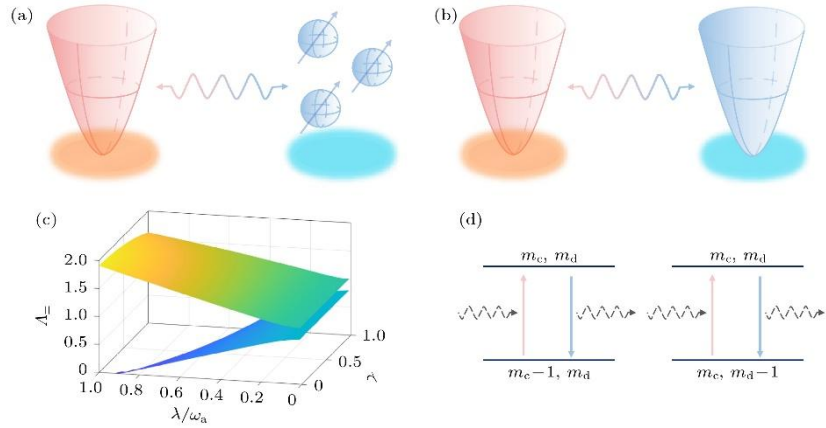

**Figure 1.** A schematic description of (a) anisotropic Dicke model and (b) two-coupled-oscillator model, of which these quantum components, i.e., qubits and photons, individually interact with bosonic thermal reservoirs. (c) Two eigenmodes of two-coupled-oscillator Hamiltonian at Eq. (3) with $\omega_a = 1$, $\varepsilon = 0.8\omega_a$. (d) Incoherent energy exchange processes between the two-coupled-oscillator system in the eigen-basis and the thermal reservoirs.

Where $\gamma$ denotes the anisotropy coefficient. When $\gamma=1$, the anisotropic Dicke model reduces to the standard quantum Dicke model in equation (1) $\hat{H}_{\text{Dicke}}$.

In the thermodynamic limit $(N_s \to \infty)$, the Holstein-Primakoff transformations $\hat{J}_+ =$

$\hat{b}^\dagger \sqrt{N_s - \hat{b}^\dagger \hat{b}}$, $\hat{J}_- = \sqrt{N_s - \hat{b}^\dagger \hat{b}}\,\hat{b}$, and $\hat{J}_z = (\hat{b}^\dagger \hat{b} - N_s/2)$ can be introduced to represent the angular momentum operator. The $\hat{b}^\dagger, \hat{b}$ is the effective Bose operator describing the angular momentum excitation. When the excitation is weak ($\langle \hat{b}^\dagger \hat{b}\rangle \ll N_s$), the angular momentum can be reduced to $\hat{J}_+ = \sqrt{N_s}\,\hat{b}^\dagger$, $\hat{J}_- = \sqrt{N_s}\,\hat{b}$ and $\hat{J}_z = \hat{b}^\dagger \hat{b} - N_s/2$. Subsequently, the anisotropic Dicke model in Eq. (2) reduces to the coupled harmonic oscillator model in Fig. 1(b). In this case, the Hamiltonian is approximately

$$\hat{H}'_s = \varepsilon\left(\hat{b}^\dagger \hat{b} - N_s/2\right) + \omega_a \hat{a}^\dagger \hat{a} + \lambda[(\hat{a}^\dagger \hat{b} + \hat{a}\hat{b}^\dagger) + \gamma(\hat{a}^\dagger \hat{b}^\dagger + \hat{a}\hat{b})]. \tag{3}$$

However, when the system is $\langle \hat{b}^\dagger \hat{b}\rangle \sim N_s$ under the condition of macroscopic excitation, the equation (3) will fail.

In general, it is difficult to solve the eigenproblem of the equation (3) by analytical methods for arbitrary anisotropic coefficients $\gamma$. However, the eigenmode $\Lambda_\pm$ of the $\hat{H}'_s$ can be obtained numerically (see 附录 A). In this way, we can clearly see the reasonable range of photon-qubit coupling strength and anisotropy parameter corresponding to the eigensolution of equation (3). Fig. 1(c) numerically shows the effect of $\lambda$ and $\gamma$ on the eigenmodes. It can be seen that with the increase of the photon-qubit coupling strength $\lambda$, the frequency of the high frequency branch will be enhanced, while the frequency of the low frequency branch will be suppressed. In addition, enhancing the anisotropy coefficient $\gamma$ will reduce the effective parameter range of the coupled harmonic oscillator model. In the $\gamma=0$ limit, two modified bosonic operators $\hat{c} = \cos\frac{\theta}{2}\hat{a} + \sin\frac{\theta}{2}\hat{b}$ and $\hat{d} = -\sin\frac{\theta}{2}\hat{a} + \cos\frac{\theta}{2}\hat{b}$. Using the Bogoliubov transformation to diagonalize $\hat{H}'_s(\gamma = 0)$, we thus obtain

$$\hat{H}'_s(\gamma = 0) = \Lambda_c \hat{c}^\dagger \hat{c} + \Lambda_d \hat{d}^\dagger \hat{d}. \tag{4}$$

Here, $\tan\theta = 2\lambda/(\omega_a - \varepsilon)$, the eigenfrequencies are $\Lambda_c = (\omega_a + \varepsilon)/2 + \sqrt{(\omega_a - \varepsilon)^2/4 + \lambda^2}$ and $\Lambda_d = (\omega_a + \varepsilon)/2 - \sqrt{(\omega_a - \varepsilon)^2/4 + \lambda^2}$.

When $\gamma=1$, $\hat{H}'_s(\gamma=1)$ can be diagonalized as[61]

$$\hat{H}'_s(\gamma=1) = \Lambda_A \hat{A}^\dagger \hat{A} + \Lambda_B \hat{B}^\dagger \hat{B}, \qquad (5)$$

Where the corrected boson operator becomes

$$\hat{A} = \frac{1}{2}\left\{ \frac{\sin\theta'}{\sqrt{\omega_a \Lambda_A}}[(\Lambda_A - \omega_a)\hat{a}^\dagger + (\Lambda_A + \omega_a)\hat{a}] + \frac{\cos\theta'}{\sqrt{\varepsilon \Lambda_A}} \right\},$$

$$\hat{B} = \frac{1}{2}\left\{ \frac{\cos\theta'}{\sqrt{\omega_a \Lambda_B}}[(\Lambda_B + \omega_a)\hat{a}^\dagger + (\Lambda_B - \omega_a)\hat{a}] \right.$$
$$\left. - \frac{\sin\theta'}{\sqrt{\varepsilon \Lambda_B}}[(\Lambda_B + \varepsilon)\hat{b}^\dagger + (\Lambda_B - \varepsilon)\hat{b}] \right\},$$

$$\tan\theta' = 4\lambda\sqrt{\omega_a \varepsilon}/(\varepsilon^2 - \omega_a^2),$$

As well as the eigenfrequencies are

$$\Lambda_A = \sqrt{[\omega_a^2 + \varepsilon^2 + \sqrt{(\varepsilon^2 - \omega_a^2)^2 + 16\lambda^2 \omega_a \varepsilon}]/2},$$

$$\Lambda_B = \sqrt{[\omega_a^2 + \varepsilon^2 - \sqrt{(\varepsilon^2 - \omega_a^2)^2 + 16\lambda^2 \omega_a \varepsilon}]/2}.$$

**2.2 Quantum master equation**

In reality, a quantum system must interact with the external environment. In this paper, we consider that photons and qubits interact with bosonic reservoirs respectively. The Hamiltonian of the nonequilibrium anisotropic Dicke model is expressed as

$$\hat{H} = \hat{H}_s + \sum_{\mu=q,r}(\hat{H}_{B,\mu} + \hat{V}_\mu). \qquad (6)$$

Where $\hat{H}_{B,\mu} = \sum_k \omega_{k\mu}\hat{b}^\dagger_{k\mu}\hat{b}_{k\mu}$ represents the boson reservoir $\mu$, and $\hat{b}^\dagger_{k\mu}(\hat{b}_{k\mu})$ represents the creation (annihilation) of a photon with frequency $\omega_{k,\mu}$ and momentum $k$ in the boson reservoir $\mu$. The interaction between the qubit and the q-reservoir is denoted by $\hat{V}_q = \frac{2\hat{J}_x}{\sqrt{N_s}}\sum_k (g_{kq}\hat{b}^\dagger_{kq} + g^*_{kq}\hat{b}_{kq})$, while the interaction between the photon and the r-reservoir is given by $\hat{V}_r = \sum_k (g_{kr}\hat{b}_{kr} + g^*_{kr}\hat{b}_{kr})(\hat{a}^\dagger + \hat{a})$. Here, $g_{k\mu}$ denotes the coupling strength and $\gamma_\mu(\omega) = 2\pi \sum_k |g_{k,\mu}|^2 \delta(\omega - \omega_k)$ is the spectral function of the $\mu$ reservoir. In this paper, an Ohmic spectral function, $\gamma_\mu(\omega) = \alpha_\mu \omega \exp(-|\omega|/\omega_c)$ [38], is used, where $\alpha_\mu$ is the dissipation strength and $\omega_c$ is the cutoff frequency of the reservoir.

The interaction $\hat{V}_\mu$ between the system and the reservoir is assumed to be weak, which can be treated perturbatively. Under the Born-Markov approximation, the dressed-state master equation (DME)[60] can be obtained as

$$\frac{\partial \hat{\rho}_s}{\partial t} = -i[\hat{H}_s, \hat{\rho}_s] + \sum_{k>j,\mu} \{\Gamma^{kj}_{\mu,-} \mathcal{D}[|\phi_j\rangle\langle\phi_k|, \hat{\rho}_s] + \Gamma^{kj}_{\mu,+} \mathcal{D}[|\phi_k\rangle\langle\phi_j|, \hat{\rho}_s]\}. \quad (7)$$

The incoherent transition rates between different eigenlevels in DME are concretely

$$\Gamma^{kj}_{\mu,+} = \gamma_\mu(\Delta_{kj}) n_\mu(\Delta_{kj}) |\langle\phi_k|\hat{A}_\mu|\phi_j\rangle|^2, \quad (8a)$$

$$\Gamma^{kj}_{\mu,-} = \gamma_\mu(\Delta_{kj})[1 + n_\mu(\Delta_{kj})] |\langle\phi_k|\hat{A}_\mu|\phi_j\rangle|^2, \quad (8b)$$

Where $\gamma_\mu(\Delta_{kj})$ stands for the spectral function of the $\mu$ reservoir, $\Delta_{kj} = E_k - E_j$ is the energy level difference of the eigenvalues in the $\hat{H}_s|\phi_k\rangle = E_k|\phi_k\rangle$, and $n_\mu(\Delta_{kj}) = 1/[\exp(\Delta_{kj}/k_B T_\mu) - 1]$ is the Bose-Einstein distribution function. In this distribution function, $k_B$ is Boltzmann's constant and $T_\mu$ is the temperature of the $\mu$ reservoir. The transition operators of the qubit and photon are $\hat{A}_q = 2\hat{J}_x/\sqrt{N_s}$ and $\hat{A}_r = (\hat{a}^\dagger + \hat{a})$, respectively, and the dissipative superoperator is $\mathcal{D}[\hat{O}, \hat{\rho}_s] = (2\hat{O}\hat{\rho}_s\hat{O}^\dagger - \hat{O}^\dagger\hat{O}\hat{\rho}_s - \hat{\rho}_s\hat{O}^\dagger\hat{O})/2$. $\Gamma^{kj}_{\mu,+}$ represents the rate at which the $\mu$ reservoir participates in the upward transition from level $j$ to level $k$, while $\Gamma^{kj}_{\mu,-}$ represents the rate at which the $\mu$ reservoir participates in the downward transition from level $k$ to level $j$.

While under the effective coupled harmonic oscillator model, the non-zero transition rate at $\gamma=0$ is expressed as

$$\Gamma_{\mu,\pm}^{mm'} = \Gamma_{\mu,\pm}^{m_c-1}(\Lambda_c)\delta_{m_c,m'_c+1}\delta_{m_d,m'_d} \\ + \Gamma_{\mu,\pm}^{m_d-1}(\Lambda_d)\delta_{m_c,m'_c}\delta_{m_d,m'_d+1}, \quad (9)$$

Where $\mu = \text{r, q}$, $\boldsymbol{m} = (m_c, m_d)$ and $\boldsymbol{m'} = (m'_c, m'_d)$, where $m_c, m'_c$ and $m_d, m'_d$ are the occupation numbers of the c-harmonic oscillator and the d-harmonic oscillator, respectively (4) where the respective transition rates are expressed as

$$\Gamma_{r,\pm}^{m_c-1}(\Lambda_c) = \gamma_r(\pm\Lambda_c)n_r(\pm\Lambda_c)\cos^2\frac{\theta}{2}m_c, \quad (10a)$$

$$\Gamma_{r,\pm}^{m_d-1}(\Lambda_d) = \gamma_r(\pm\Lambda_d)n_r(\pm\Lambda_d)\sin^2\frac{\theta}{2}m_d, \quad (10b)$$

$$\Gamma_{q,\pm}^{m_c-1}(\Lambda_c) = \gamma_q(\pm\Lambda_c)n_q(\pm\Lambda_c)\sin^2\frac{\theta}{2}m_c, \quad (10c)$$

$$\Gamma_{q,\pm}^{m_d-1}(\Lambda_d) = \gamma_q(\pm\Lambda_d)n_q(\pm\Lambda_d)\cos^2\frac{\theta}{2}m_d. \quad (10d)$$

These transition processes are described in the Fig. 1(d). It can be found that one kind of transition process is dominated by the c-harmonic oscillator, and the other is dominated by the d-harmonic oscillator. In contrast, the non-zero transition rate at $\gamma=1$ becomes

$$\Gamma_{r,\pm}^{m_a-1}(\Lambda_c) = \gamma_r(\pm\Lambda_A)n_r(\pm\Lambda_A)\cos^2\left(\frac{\theta'}{2}\right)\frac{\omega_a}{\Lambda_A}m_a, \quad (11a)$$

$$\Gamma_{r,\pm}^{m_b-1}(\Lambda_B) = \gamma_r(\pm\Lambda_B)n_r(\pm\Lambda_B)\sin^2\left(\frac{\theta'}{2}\right)\frac{\omega_a}{\Lambda_B}m_b, \quad (11b)$$

$$\Gamma_{q,\pm}^{m_a-1}(\Lambda_A) = \gamma_q(\pm\Lambda_A)n_q(\pm\Lambda_A)\sin^2\left(\frac{\theta'}{2}\right)\frac{\varepsilon}{\Lambda_A}m_a, \quad (11c)$$

$$\Gamma_{q,\pm}^{m_b-1}(\Lambda_B) = \gamma_q(\pm\Lambda_B)n_q(\pm\Lambda_B)\cos^2\left(\frac{\theta'}{2}\right)\frac{\varepsilon}{\Lambda_B}m_b. \quad (11d)$$

Similarly, the above transition process can also be described by the Fig. 1(d) framework. These different expressions of the incoherent transition rate will significantly change the dynamic and steady-state behavior of the coupled harmonic oscillator model. The anisotropic Dicke model exhibits a finite quantum heat flow in the steady state in the presence of a finite temperature difference $T_r \neq T_q$. In the subsequent calculation, in order to obtain the convergence value of the heat flow in the range shown in the figure, the cut-off number of photons is set to $N_{tr}^a = 30$. It is relatively easy to further increase the truncation number in the calculation, but the heat flow behavior is not changed.

**3. Steady state heat transport**

In this part, the effect of anisotropic photon-qubit interaction on the steady-state heat flow of the anisotropic Dicke model is first studied. Based on the microscopic transition process exhibited by equation (7) DME, the general expression of the steady-state heat flow in a $\mu$ reservoir is

$$J_\mu = \sum_{k>n} \Delta_{kn} (\Gamma_{\mu,-}^{k,n} P_k - \Gamma_{\mu,+}^{k,n} P_n), \quad (12)$$

Where $\Delta_{kn} = E_k - E_n$ denotes the energy gap and $P_k = \langle \phi_k | \hat{\rho}_s | \phi_k \rangle$ is the probability distribution in the steady state. The effect of anisotropic photon-qubit coupling strength on the steady state heat flow is analyzed in the finite size and thermodynamic limits, respectively. The effect of thermal rectification in the anisotropic Dicke model is also studied.

**3.1 Steady state heat flow**

One of the main motivations of this paper is to analyze the effect of anisotropic parameters on the steady state heat flux through Fig. 2. The heat flow behavior of the anisotropic Dicke model in the one-qubit limit is shown Fig. 2(a). In the photon-qubit coupling, the eigenenergy spectrum and eigenstates of the anisotropic QRM system can be approximately obtained by the JC model. The counter-rotating wave term does not provide an efficient energy exchange process. Therefore, the heat flow is insensitive to the anisotropy coefficient $\gamma$. The heat flow increases synchronously with the increase of the coupling strength. However, the heat flow shows a non-monotonic behavior as the photon-qubit coupling strength is further increased. The specific manifestation is that the heat flow becomes weak at strong coupling. At this point, the role of the anti-rotating wave term begins to become apparent. Multiphoton scattering makes incoherent transitions difficult[24]. On the whole, the heat flow is first enhanced at weak coupling and suppressed at strong coupling. Therefore, the anisotropic photon-qubit strong coupling is crucial to regulate the heat flow of the anisotropic QRM. The effects of photon-qubit coupling strength and anisotropy coefficient on heat flow in the case of multi-qubit ($N_s$=2,4,6) are presented in Fig. 2(b) —(d). As the number of qubits increases, the heat flow peak becomes higher at moderate coupling strengths. However, the heat flow valley decreases significantly at large anisotropy and strong photon-qubit coupling, such as the red line corresponding to $\gamma$=1 in Fig. 2(d). Therefore, multiple qubits amplify the heat flow change signal under the strong photon-qubit coupling.

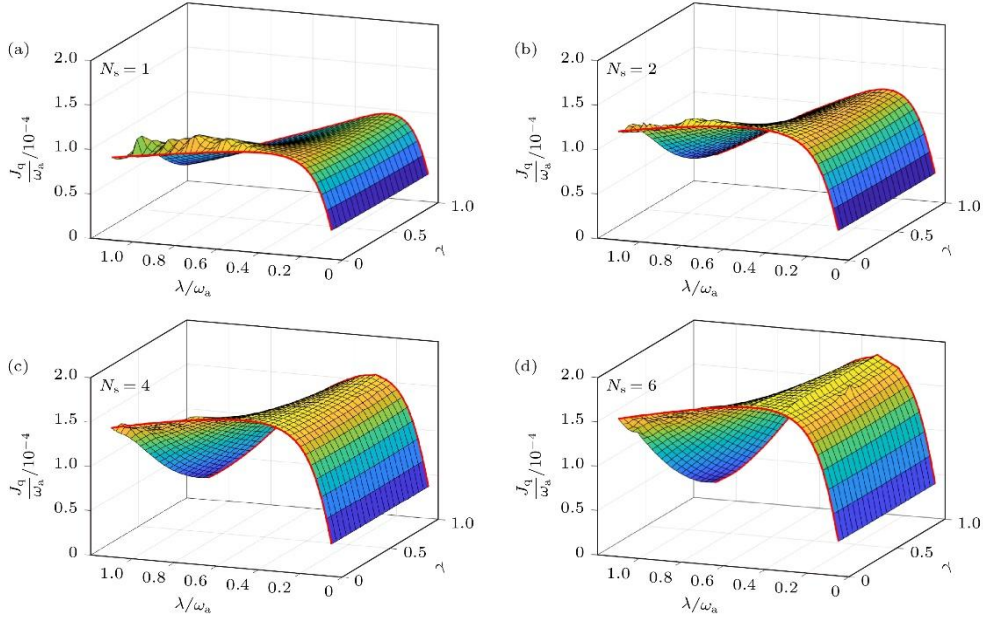

**Figure 2.** Influences of qubit-photon coupling strength $\lambda$ and anisotropic factor $\gamma$ on steady state heat flow $J_q$ in (a) $N_s = 1$, and finite numbers of qubits (b) $N_s = 2$, (c) $N_s = 4$, and (d) $N_s = 6$. The redlines denote heat flows at $\gamma = 0$ and $\gamma = 1$ limiting cases. System parameters are given by $\omega_a = 1$, $\varepsilon = 0.8\omega_a$, $\omega_c = 20\omega_a$, $\alpha_r = \alpha_q = 0.001\omega_a$, $T_r = 1.2\omega_a$, and $T_q = 0.6\omega_a$.

In addition, the analytical form of the heat flow in the thermodynamic limit of the coupled harmonic oscillator model of Eq. (3) is analyzed. When $\gamma=0$, the Hamiltonian of the mixed quantum system is described by the equation (4). While the incoherent transition rate associated with the reservoir is exhibited in the equation (9). Therefore, the particle number distribution in the steady state is

$$P_{m_c,m_d} = \frac{e^{-\beta_{eff}^c m_c \Lambda_c}}{1+n_c} \times \frac{e^{-\beta_{eff}^d m_d \Lambda_d}}{1+n_d}, \tag{13}$$

Where the Bose-Einstein distribution function is expressed as $n_\mu = 1/[\exp(\beta_{eff}^\mu \Lambda_\mu) - 1]$ ($\mu = c, d$) The effective inverse temperature $\beta_{eff}^\mu$ can be obtained from the following formula:

$$= \frac{\exp(-\beta_{\text{eff}}^c \Lambda_c)}{\sum_{x=r,q} \gamma_x(\Lambda_c) n_x(\Lambda_c) \cos^2(\theta_x/2)}{\sum_{x=r,q} \gamma_x(-\Lambda_c) n_x(-\Lambda_c) \cos^2(\theta_x/2)}, \quad (14a)$$

$$= \frac{\exp(-\beta_{\text{eff}}^d \Lambda_d)}{\sum_{x=r,q} \gamma_x(\Lambda_d) n_x(\Lambda_d) \sin^2(\theta_x/2)}{\sum_{x=r,q} \gamma_x(-\Lambda_d) n_x(-\Lambda_d) \sin^2(\theta_x/2)}, \quad (14b)$$

Where $\theta_r = \theta$, $\theta_q = \pi - \theta$, $\tan\theta = 4\lambda\sqrt{\omega_a \varepsilon}/(\varepsilon^2 - \omega_a^2)$, and $\gamma_x(-\Lambda_\mu) n_x(-\Lambda_\mu) = \gamma_x(\Lambda_\mu)[1 + n_x(\Lambda_\mu)]$ satisfies the detailed balance relation. According to the definition of heat flow in equation (12), the analytical expression of heat flow into the q-heat reservoir is

$$J_q^{\gamma=0} = \Lambda_c \sin^2(\theta/2)\gamma_q(\Lambda_c)[(1 + n_q(\Lambda_c))n_c \\ - n_q(\Lambda_c)(1 + n_c)] + \Lambda_d \cos^2(\theta/2)\gamma_q(\Lambda_d) \\ \times [(1 + n_q(\Lambda_d))n_d - n_q(\Lambda_d)(1 + n_d)]. \quad (15)$$

When $\gamma=1$, the Hamiltonian of the hybrid quantum system is Eq. (5). In this case, Eq. (11) gives the corresponding incoherent transition rate. The corresponding steady state of the system is

$$P_{m_A, m_B} = \frac{e^{-\beta_{\text{eff}}^A m_A \Lambda_A}}{1 + n_A} \times \frac{e^{-\beta_{\text{eff}}^B m_B \Lambda_B}}{1 + n_B}, \quad (16)$$

Where the Bose-Einstein distribution function is expressed as $n_\mu = 1/[\exp(\beta_{\text{eff}}^\mu \Lambda_\mu) - 1]$ ($\mu = A, B$) and the effective inverse temperature can be obtained by

$$\exp(-\beta_{\text{eff}}^A \Lambda_A) = \frac{\gamma_r(\Lambda_A) n_r(\Lambda_A) \cos^2(\theta'/2)(\omega_a/\Lambda_A) + \gamma_q(\Lambda_A) n_q(\Lambda_A) \sin^2(\theta'/2)(\varepsilon/\Lambda_A)}{\gamma_r(-\Lambda_A) n_r(-\Lambda_A) \cos^2(\theta'/2)(\omega_a/\Lambda_A) + \gamma_q(-\Lambda_A) n_q(-\Lambda_A) \sin^2(\theta'/2)(\varepsilon/\Lambda_A)}, \quad (17a)$$

$$\exp(-\beta_{\text{eff}}^B \Lambda_B) = \frac{\gamma_r(\Lambda_B) n_r(\Lambda_B) \sin^2(\theta'/2)(\omega_a/\Lambda_B) + \gamma_q(\Lambda_B) n_q(\Lambda_B) \cos^2(\theta'/2)(\varepsilon/\Lambda_B)}{\gamma_r(-\Lambda_B) n_r(-\Lambda_B) \sin^2(\theta'/2)(\omega_a/\Lambda_B) + \gamma_q(-\Lambda_B) n_q(-\Lambda_B) \cos^2(\theta'/2)(\varepsilon/\Lambda_B)}, \quad (17b)$$

Where $\tan\theta' = 4\lambda\sqrt{\omega_a \varepsilon}/(\varepsilon^2 - \omega_a^2)$. Therefore, the heat flow into the q-heat reservoir is

$$J_q^{\gamma=1} = \varepsilon\left\{\sin^2\frac{\theta'}{2}\gamma_q(\Lambda_A)[(1+n_q(\Lambda_A))n_A - n_q(\Lambda_A)(1+n_A)] + \cos^2\frac{\theta'}{2}\gamma_q(\Lambda_B)[(1+n_q(\Lambda_B))n_B - n_q(\Lambda_B)(1+n_B)]\right\}. \quad (18)$$

The [Fig. 3(a)](#) describes the effect of different coupling strengths on the steady-state heat flux of the nonequilibrium coupled harmonic oscillator model with different anisotropy coefficients. The $J_q$ is almost constant with anisotropy coefficient at weak coupling. This indicates that the microscopic transition process is dominated by the rotating wave term $\lambda(\hat{a}^\dagger \hat{b} + \hat{b}^\dagger \hat{a})$ of the coupled harmonic oscillator model. At strong coupling, the anti-rotating wave term begins to play an important role. Especially at large anisotropy, the strong photon-qubit coupling will significantly suppress the heat flow. This is consistent with the result in the finite-size anisotropic Dicke model. The [Fig. 3(b)](#) show that when the coupling strength of the coupled harmonic oscillator is moderate, the heat flow will increase with the increase of the anisotropy coefficient $\gamma$. In addition, based on Eq. ([15](#)) and Eq. ([18](#)), it is found that the heat flow exhibits co-tunneling transport characteristics at $\gamma=0$ and $\gamma=1$. These heat flow solutions also describe the upper limit of the heat flow in the finite-size anisotropic Dicke model as the number of bits tends to infinity. Therefore, the anisotropic photon-qubit strong coupling can effectively modulate the behavior of the heat flow. The nonequilibrium coupled harmonic oscillator model can provide a physical picture of the finite-size anisotropic Dicke model in the thermodynamic limit.

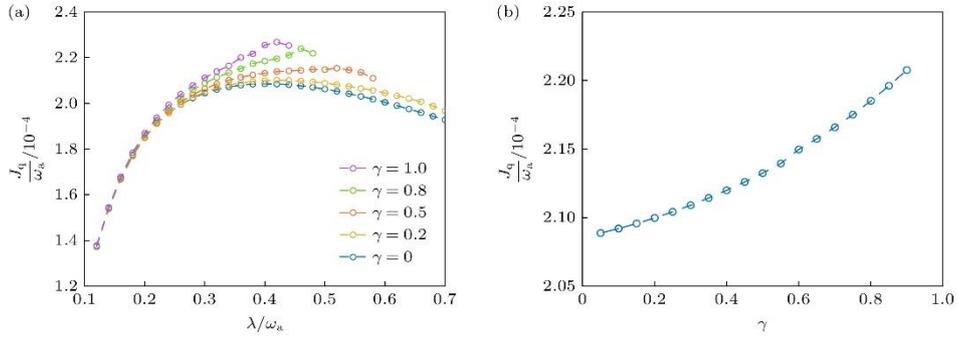

**Figure 3.** (a) Steady-state heat flow of the nonequilibrium two-coupled-oscillator model by tuning the qubit-photon interaction strength with various anisotropic factors; (b) the influence of anisotropic factor on the heat flow at $\lambda/\omega_a = 0.4$. Other system parameters are given by $\omega_a = 1$, $\varepsilon = 0.8\omega_a$, $\omega_c = 20\omega_a$, $\alpha_r = \alpha_q = 0.001\omega_a$, $T_r = 1.2\omega_a$, and $T_q = 0.6\omega_a$.

### 3.2 Thermal rectification

Thermal rectification is widely recognized as one of the key effects characterizing non-equilibrium thermal devices[62]. Similar to electronic devices, thermal devices also exhibit nonreciprocal behavior based on two-terminal setup systems. The thermal rectification effect characterizes the degree of difference between the heat flow in one direction and the heat flow in the opposite direction obtained by only exchanging the temperatures of the two reservoirs. This description of thermal rectification was originally proposed by Li et al.[63,64] in a classical nonlinear harmonic oscillator lattice. The thermal rectification factor can be expressed by the following equation:

$$\mathcal{R} = \frac{|J_q(\Delta T) + J_q(-\Delta T)|}{\max\{|J_q(\Delta T)|, |J_q(-\Delta T)|\}}, \quad (19)$$

Where $J_q(\pm \Delta T)$ represents the heat flow under biased conditions of $T_r = T_0 + \Delta T/2$ and $T_q = T_0 - \Delta T/2$. When the nonequilibrium quantum system has reciprocity $J_q(\Delta T) = -J_q(-\Delta T)$, the rectification factor $\mathcal{R}$ approaches 0. On the contrary, when $\mathcal{R}$ approaches 1, the quantum system exhibits high nonreciprocity $|J_q(\Delta T)| \gg |J_q(-\Delta T)|$.

It can be found that the thermal rectification coefficient is closely related to the temperature difference through the (19) formula. The effects of temperature bias $\Delta T$ and photon-qubit coupling strength $\lambda$ on the rectification factor $\mathcal{R}$ are investigated. The rectification factor results at different anisotropy coefficients are shown in Fig. 4(a) —(c). It can be seen that the rectification factor R increases monotonically with the increase of temperature deviation. When $\gamma$ tends to 1, the maximum value of R is close to 0.5. However, the increase of photon-qubit coupling strength leads to a non-monotonic behavior of $\mathcal{R}$. $\mathcal{R}$ shows an optimal peak at both weak and moderate coupling. There is a sharp valley between these two peaks that tends to 0. At this time, the heat flow shows the reciprocal characteristics. In addition, a large anisotropy coefficient can significantly improve the $\mathcal{R}$ at moderate photon-qubit coupling (for example, the maximum value of $\mathcal{R}$ in Fig. 4(c) is about 0.45). The maximum value of $\mathcal{R}$ is also studied as a function of the photon-qubit coupling strength and the anisotropy coefficient. Fig. 4(d)-(f) show the behavior of the rectification factor for different numbers of qubits. By modulating the larger anisotropy coefficient, it is found that the $\max_{\Delta T}\{R\}$ has two maximum peaks in the moderate and strong coupling regions, and the extreme values are close to each other 0.5. Therefore, large temperature deviation, large anisotropy coefficient and non-weak photon-qubit coupling are helpful to achieve significant thermal rectification effect.

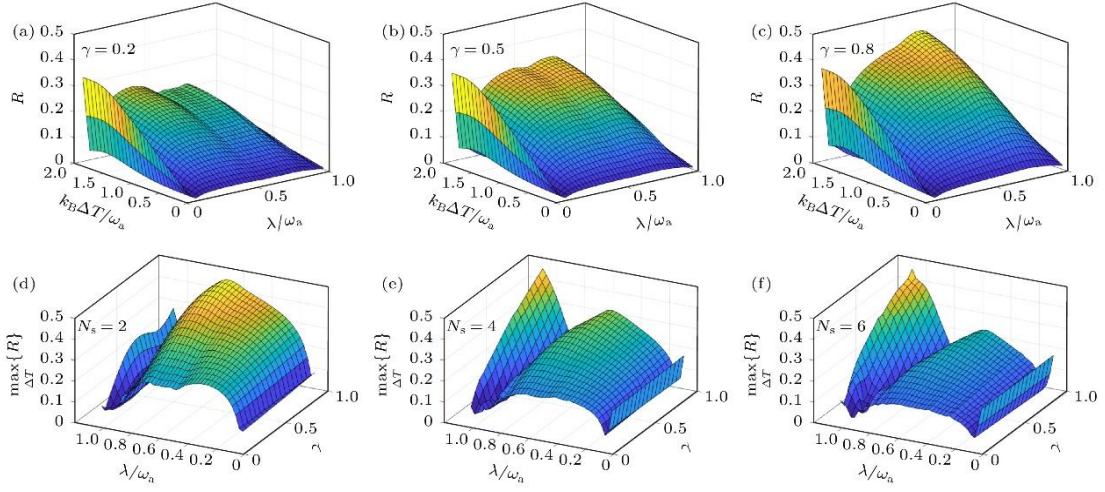

**Figure 4.** Thermal rectification factor $\mathcal{R}$ by tuning qubit-photon coupling strength $\lambda$ and temperature bias $\Delta T$ ($T_r = T_0 + \Delta T/2$, $T_q = T_0 - \Delta T/2$, and $T_0 = \omega_a$) with $N_s = 2$ at (a) $\gamma = 0.2$, (b) $\gamma = 0.5$, and (c) $\gamma = 0.8$. Maximal value of $\mathcal{R}$ by searching over the temperature bias as a function of $\lambda$ and $\gamma$ with (d) $N_s = 2$, (e) $N_s = 4$, and (f) $N_s = 6$. Other system parameters are given by $\omega_a = 1$, $\varepsilon = 0.8\omega_a$, $\omega_c = 20\omega_a$, and $\alpha_r = \alpha_q = 0.001\omega_a$.

## 4. Conclusion

Based on the master equation of quantum dressed States, the behavior of quantum heat flow and the effect of heat rectification in the nonequilibrium anisotropic Dicke model are studied. The photon-qubit strong coupling is properly treated by introducing a quantum dressed-state master equation into the intrinsic picture of the Dicke model. In this paper, the effects of photon-qubit coupling strength and anisotropy on the steady state heat flow are studied. The results show that the anisotropic photon-qubit strong coupling can suppress the heat flow of the anisotropic Dicke model; Anisotropic interaction at moderate coupling strength enhances the heat flow behavior. And the increase of the number of qubits will amplify the heat flow signal under the strong photon-qubit coupling. Therefore, the anisotropic photon-qubit interaction can significantly modulate the behavior of the heat flow. In addition, based on the coupled harmonic oscillator model in the thermodynamic limit approximation, the expressions of the heat flow at $\gamma=0$ and $\gamma=1$ are obtained analytically. These two analytical expressions show that the corresponding heat flux exhibits a cotunneling transport picture and provide a theoretical upper limit for the heat flux of the anisotropic Dicke model at finite size. The thermal rectification effect is also studied, and it is found that large temperature deviation, large anisotropy and non-weak photon-qubit coupling can achieve significant thermal rectification effect. This study provides a theoretical basis for the control of non-equilibrium thermal energy transport and quantum thermal device effects in anisotropic light-matter interaction systems.

**Appendix A Eigenmodes of the A Bose operator.**

The dynamic Bose operator can be obtained based on $\hat{H}'_s$ in Eq. (3)

$$d\hat{a}^\dagger/dt = i(\omega_a \hat{a}^\dagger + \lambda \hat{b}^\dagger + \lambda\gamma \hat{b}), \quad (A1a)$$

$$d\hat{a}/dt = -i(\omega_a \hat{a} + \lambda \hat{b} + \lambda\gamma \hat{b}^\dagger), \quad (A1b)$$

$$d\hat{b}^\dagger/dt = i(\varepsilon \hat{b}^\dagger + \lambda \hat{a}^\dagger + \lambda\gamma \hat{a}), \quad (A1c)$$

$$d\hat{b}/dt = -i(\varepsilon \hat{b} + \lambda \hat{a} + \lambda\gamma \hat{a}^\dagger). \quad (A1d)$$

If $\boldsymbol{O} = [\hat{a}^\dagger; \hat{b}^\dagger; \hat{a}; \hat{b}]$ is defined, the dynamic equation can be expressed as $d\boldsymbol{O}/dt = i\boldsymbol{H}\boldsymbol{O}$. Where the block form of the evolution matrix $\boldsymbol{H}$ is $\boldsymbol{H} = [\boldsymbol{A}, \boldsymbol{B}; -\boldsymbol{B}, -\boldsymbol{A}]$, the matrix $\boldsymbol{A} = [\omega_a, \lambda; \lambda, \varepsilon]$, the matrix $\boldsymbol{B} = [0, \lambda\gamma; \lambda\gamma, 0]$. Interestingly, the eigenvalues of $\boldsymbol{H}$ always come in pairs. For example, if $\boldsymbol{H}[u,v]^T = \Lambda[u,v]^T (\Lambda > 0)$, then directly $\boldsymbol{H}[v,u]^T = -\Lambda[v,u]^T$. Through dynamic analysis, the two positive real eigenvalues $\Lambda_\pm$ of $\boldsymbol{H}$ are the eigenmodes of $\hat{H}_s$, and $\hat{H}_s$ can be expressed as $\hat{H}_s = \Lambda_+ \hat{c}^\dagger \hat{c} + \Lambda_- \hat{d}^\dagger \hat{d}$, where $\hat{c}^\dagger = [u_+, v_+]\boldsymbol{O}$, $\hat{d}^\dagger = [u_-, v_-]\boldsymbol{O}$.

References

[1] Cohen-Tannoudji C, Dupont-Roc J, Grynberg G 1998 Atom-Photon Interactions: Basic Processes and Applications (New Jersey: Wiley-VCH) pp15–19

[2] Scully M O, Zubairy M S 1997 Quantum Optics (Cambridge: Cambridge University Press) pp193–219

[3] Haroche S, Brune M, Raimond J M 2020 Nat. Phys. 16 243

[4] Kurizki G, Bertet P, Kubo Y, Molmer K, Petrosyan D, Rabl P, Schmiedmayer J 2015 PNAS 112 3866

[5] Gonzalez-Tudela A, Reiserer A, Garcia-Ripoll J J, Garcia-Vidal F J 2024 Nat. Rev. Phys. 6 166

[6] Ronzani A, Karimi B, Senior J, Chang Y C, Peltonen J T, Chen C D, Pekola J P 2018 Nat. Phys. 14 991

[7] Pekola J P, Karimi B 2021 Rev. Mod. Phys. 93 041001

[8] Jaynes E T, Cummings F W 1963 Proc. IEEE 51 89

[9] Greentree A D, Koch J, Larson J 2013 J. Phys. B 46 220201

[10] Blais A, Girvin S M, Oliver W D 2020 Nat. Phys. 16 247

[11] Blais A, Grimsmo A L, Girvin S M, Wallraff A 2021 Rev. Mod. Phys. 93 025005

[12] Petersson K D, McFaul L W, Schroer M D, Jung M, Taylor J M, Houck A A, Petta J R 2012 Nature 490 380

[13] Lin T, Li H O, Cao G, Guo G P 2023 Chin. Phys. B 32 070307

[14] Jaako T, Garcia-Ripoll J J, Rabl P 2020 Adv. Quantum Technol. 3 1900125

[15] Cai M L, Liu Z D, Zhao W D, Wu Y K, Mei Q X, Jiang Y, He L, Zhang X, Zhou Z C, Duan L M 2021 Nat. Commun. 12 1126

[16] Rabi I I 1936 Phys. Rev. 49 324

[17] Rabi I I 1937 Phys. Rev. 51 652

[18] Braak D 2011 Phys. Rev. Lett. 107 100401

[19] Braak D, Chen Q H, Batchelor M T, Solano E 2016 J. Phys. A 49 300301

[20] Xie Q T, Cui S, Cao J P, Amico L, Fan H 2014 Phys. Rev. X 4 021046

[21] Lyu G T, Kottmann K, Plenio M B, Myung-Joong H 2024 Phys. Rev. Res. 6 033075

[22] Lu J H, Ning W, Zhu X, Wu F, Shen L T, Yang Z B, Zheng S B 2022 Phys. Rev. A 106 062616

[23] Zhu X, Lu J H, Ning W, Wu F, Shen L T, Yang Z B, Zheng S B 2023 Sci. China Phys. Mech. Astron. 66 250313

[24] Chen Z H, Che H X, Chen Z K, Wang C, Ren J 2022 Phys. Rev. Res. 4 013152

[25] Ye T, Wang C, Chen Q H 2023 Physica A 609 128364

[26] Ye T, Wang C, Chen Q H 2024 Opt. Express 32 33483

[27] Zhang Y Y, Chen X Y 2017 Phys. Rev. A 96 063821

[28] Dicke R H 1954 Phys. Rev. 93 99

[29] Kirton P, Roses M M, Keeling J, Dalla Torre E G 2019 Adv. Quantum Technol. 2 1970013

[30] Emary C, Brandes T 2003 Phys. Rev. Lett. 90 044101

[31] Lambert N, Emary C, Brandes T 2004 Phys. Rev. Lett. 92 073602

[32] Yu L X, Liang Q F, Wang L R, Zhu S Q 2014 Acta Phys. Sin. 63 134204

[33] Zhao X Q, Zhang W H, Wang H M 2024 Acta Phys. Sin. 73 160302

[34] Gyhm J Y, Safranek D, Rosa D 2022 Phys. Rev. Lett. 128 140501

[35] Dou F Q, Lu Y Q, Wang Y J, Sun J A 2022 Phys. Rev. A 106 032212

[36] Huang B Y, He Z, Chen Y 2023 Acta Phys. Sin. 72 180301

[37] Seidov S S, Mukhin S I 2024 Phys. Rev. A 109 022210

[38] Weiss U 1999 Quantum Dissipative Systems (Singapore: World Scientific) pp250, 251

[39] Chiacchio1 E I R, Nunnenkamp A, Brunelli M 2023 Phys. Rev. Lett. 131 113602

[40] Mivehvar F 2024 Phys. Rev. Lett. 132 073602

[41] Vivek G, Mondal D, Chakraborty S, Sinha S 2025 Phys. Rev. Lett. 134 113404

[42] Gong Z P, Hamazaki R, Ueda M 2018 Phys. Rev. Lett. 120 040404

[43] Jager S B, Giesen J M, Schneider I, Eggert S 2024 Phys. Rev. A 110 L010202

[44] Kirton P, Keeling J 2018 New J. Phys. 20 015009

[45] Strashko A, Kirton P, Keeling J 2018 Phys. Rev. Lett. 121 193601

[46] Das P, Bhakuni D S, Sharma A 2023 Phys. Rev. A 107 043706

[47] Chen X Y, Zhang Y Y, Chen Q H, Lin H Q 2024 Phys. Rev. A 110 063722

[48] Buijsman1 W, Gritsev V, Sprik R 2017 Phys. Rev. Lett. 118 080601

[49] Zhu X, Lu J H, Ning W, Shen L T, Wu F, Yang Z B 2024 Phys. Rev. A 109 052621

[50] Senior J, Gubaydullin A, Karimi B, Peltonen J T, Ankerhold J, Pekola J P 2020 Commun. Phys. 3 40

[51] Gubaydullin A, Thomas G, Golubev D S, Lvov D, Peltonen J T, Pekola J P 2022 Nat. Commun. 13 1552

[52] Liu Y Q, Yang Y J, Yu C S, 2023 Phys. Rev. E 107 044121

[53] Zhao X D, Xing Y, Cao J, Liu S T, Cui W X, Wang H F, 2023 npj Quantum Inf. 9 59

[54] Lu J C, Wang R Q, Ren J, Kulkarni M, Jiang J H 2019 Phys. Rev. B 99 035129

[55] Majland M, Christensen K S, Zinner N T 2020 Phys. Rev. B 101 184510

[56] Wang C, Chen H, Liao J Q 2021 Phys. Rev. A 104 033701

[57] Andolina G M, Erdman P A, Noe F, Pekola J, Schiro M 2024 Phys. Rev. Res. 6 043128

[58] Beaudoin F, Gambetta J M, Blais A 2011 Phys. Rev. A 84 043832

[59] Altintas F, Eryigit R 2013 Phys. Rev. A 87 022124

[60] Le Boite A 2020 Adv. Quantum Technol. 3 1900140

[61] Emary C, Brandes T 2003 Phys. Rev. E 67 066203

[62] Li N B, Ren J, Wang L, Zhang G, Hanggi P, Li B 2012 Rev. Mod. Phys. 84 1045

[63] Li B, Wang L, Casati G 2004 Phys. Rev. Lett. 93 184301

[64] Zhang L F, Yan Y H, Wu C Q, Wang J S, Li B W 2009 Phys. Rev. B 80 172301